\documentclass[a4paper,10pt]{article}
\usepackage[utf8]{inputenc}
\usepackage[english]{babel}
\usepackage{amsmath,amssymb}
\usepackage{geometry}
\usepackage{tabularx}
\usepackage{graphicx}
\usepackage{hyperref}
\usepackage{xcolor}
\usepackage{float}      
\usepackage{authblk}    

\title{Observational Manifestations of Primordial Objects in the Early Universe Through the Hydrogen Subordinate Lines}

\author[1]{Viktor K.~Dubrovich}
\author[2]{Yury~N.~Eroshenko}
\author[1]{Stanislav~I.~Shirokov\thanks{Correspondence: arhath.sis@yandex.ru}}
\affil[1]{Special Astrophysical Observatory, Russian Academy of Sciences, Zelenchuksky District, 369167 Nizhny Arkhyz, Russia}
\affil[2]{Institute for Nuclear Research of the Russian Academy of Sciences, 60th Anniversary of October Prospect 7a, 117312 Moscow, Russia}

\date{}

\begin{document}

\maketitle

\begin{abstract}
A new mechanism for the formation of spectral--spatial distortions in cosmic microwave background radiation near primordial massive compact objects (such as primordial black holes) at redshifts from $z \sim 1000$ to $\sim$100 is proposed. After hydrogen recombination, the radiation from these objects leads to a significant increase in the population of hydrogen subordinate levels in their surrounding environment. Consequently, this allows for the observation of a Fraunhofer-like absorption spectrum in the cosmic microwave background. Such a distortion is formed due to the temperature difference between matter and relic radiation at the corresponding epoch. Ultimately, we should observe circular absorption or emission features around the objects with small angular sizes. These subordinate hydrogen lines currently lie in the radio wavelength range. Estimates indicate that the effect under consideration is accessible for the observations with planned large radiotelescopes. The possibility of the proposed mechanism lies in the ability of an object (e.g., accretion disk around black hole) to emit few-eV photons that populate the $n=2,3,4$ and higher levels of hydrogen, enabling the required excitation.
\end{abstract}


\section{Introduction}

The study of the early Universe is of fundamental importance for our understanding of physical laws and new phenomena that are not attainable in laboratories. A large number of works on this topic, both theoretical and experimental/observational, have been carried out over the past few decades, and important results have been obtained. However, a lot of phenomena are still unexplained in this area, and new avenues have to be explored.

One of the key tasks in studying the early Universe is the investigation of the so-called ``Dark Ages''---the era after hydrogen recombination at redshifts from about $z \approx 1100$ down to $z \approx 30$, when the first stars and star systems are already forming~\cite{Metron2024}. During this period, within the framework of the standard model of the evolution of the Universe, there are no bright, readily observable sources. Employing standard methods of optical and X-ray astronomy in the Dark Ages is currently practically impossible, and the main hope lies in radio observations.

Nevertheless, there is great interest in various models that aim to answer the following question: ``Is the matter in the Universe during the Dark Ages a weakly inhomogeneous gaseous medium with a huge number of CMB photons, or do there still exist some primordial compact objects, more or less bright, similar to objects observed near us, such as quasars?'' At present, the observational capabilities of modern astronomy are limited to redshifts of $z \approx 17$. The stellar magnitudes of objects at high redshifts, their brightness temperatures, and their angular sizes are beyond the reach of current telescopes, making direct observation difficult. However, several approaches to address this problem can be proposed based on the use of indirect observations: One should observe not the objects themselves but the result of their interaction between their emitted radiation and the surrounding gas and CMB radiation. Thus, one must consider mechanisms that allow an object to be observed not through its direct radiation but through disturbances in the CMB. An example of such an effect is the mechanism outlined in~\cite{Dubrovich2021b}. This mechanism allows one to determine the presence of a gravitating object without additional energy release, based on the fact that it forms a field of velocities in the background gas and a specific temperature distribution of the radiation in the 21\,cm line around it. More generally, the 21 cm signal from neutral hydrogen during the Dark Ages can probe fundamental physics and is the target of future low-frequency space interferometers~\cite{Koopmans2021}.

In this article, we propose a new approach to observe massive compact objects in the early Universe based on the absorption or emission of CMB radiation at highly excited levels of hydrogen that occur near such objects under the influence of their radiation. For the case of relatively close space objects, the possibility of radio observations of subordinate lines (radio recombination lines) was proposed by van de Hulst in 1945~\cite{Hul45} and was considered in detail by N.S.~Kardashev in 1959~\cite{Kar59}; then, this effect was discovered in observations using radio telescopes. As a development of this research method, we propose looking for similar hydrogen lines from objects in the early Universe at the pre-galactic stage of its evolution.

Highly excited levels of hydrogen atoms in the early Universe have been considered in a number of papers, but these were considered in the case of a homogeneous medium (e.g., ref. \cite{ChlRubSun07} considered levels up to $n\sim100$ and their effect on the course of the recombination process). In works~\cite{Liuetal23-1,Liuetal23-2}, levels up to $n\sim200$ were considered, the population of which was maintained by additional sources of energy release, for example, due to the decay of dark matter particles, and the distortion of the spectrum of cosmic microwave background radiation caused by this process was investigated. However, in works~\cite{Liuetal23-1,Liuetal23-2}, only uniformly distributed sources of energy release were considered, for example, the homogeneous cosmological background of decaying dark matter particles. In this paper, we consider a fundamentally different situation in which there are massive isolated radiation sources. Such sources can change the population of levels in hydrogen in the space around them. The most interesting effect occurs with excitations to levels with a large $n$, since the corresponding transitions can manifest themselves as subordinate radio lines. The absorption or emission regions most likely have the appearance of concentric rings around massive objects. The discovery of such rings would be of exceptional interest to cosmology, as it would reveal the existence of rare exotic objects with high mass in the Universe.

Several possible variants of the formation of massive compact cosmological objects radiating energy in the early Universe have been proposed. These objects could be single, accreting, supermassive primordial black holes (PBHs) \cite{ZelNov67,Haw71,Car81,Carr2024,Bicknell1979,Volonteri2021,Davies2011,Lupi2014,Belotsky2019,Ziparo2025,Cai2024,Dayal2025,Sobrinho2024,Li2025,Hooper2024,LiuBromm2025,Mould2025,Prole2025}; a cluster of low-mass evaporating PBHs~\cite{Beletal11,Beletal14} (the formation of such clusters was proposed in~\cite{RubKhlSak00,RubSakKhl01}); wormholes through which matter is ejected from the hot regions of our or other universes; or massive clumps of dark matter from annihilating or decaying particles. Energy emission is also possible during the decay of cosmic string loops. Silk and Dolgov also proposed a model for the formation of massive objects from regions with baryon charge fluctuations, including those with an excess of antimatter~\cite{DolSil93}. In this paper, we consider the possibility of detecting these or other massive objects by observing the subordinate lines of hydrogen atoms in the radio band. The detection of hydrogen recombination lines produced by growing black holes was proposed in the work of \cite{VasSetShc18}, but only for relatively close objects at $z\sim8.5$--$16.5$.

Due to Thomson scattering, the temperature of the gas at redshifts $z>150$ is very close to the CMB temperature. When $z$ changes from 200 to 1250, the relative temperature difference changes from $\sim$0.1 to $\sim$$10^{-6}$ (see Figure~5 in~\cite{Kho11}). This fact leads to a significant suppression of the effect considered in this paper at large $z$ in the case of the search for absorption. Nevertheless, microwave spectroscopy already has very high sensitivity and continues to develop; therefore, even taking into account this suppression, the effects of subordinate transitions under consideration may be registered in the near future.


\section{Physical Mechanism of Population Enhancement}
\label{sec:mechanism}

\subsection{Basic Equations and Level Populations}

In this work, we draw attention to an interesting feature of highly excited hydrogen lines at high redshifts. The population of such levels is determined by the combined action of ultraviolet radiation from the central source in the near Lyman-$\alpha$ band and the background CMB radiation in subordinate transitions. The main problem is that the radiation power in the lines of the Lyman series, particularly Lyman-$\alpha$, in the case of equilibrium CMB radiation turns out to be very small. The Boltzmann factor in the Wien tail of the Planckian spectrum reaches a value of approximately $10^{-20}$ at $z = 1000$. This implies a giant suppression of the probability of excitation of the second, third, and higher levels of hydrogen.

It turns out that the low intensity of the background radiation can be overcome by additional radiation in the Lyman series from point sources like quasars. Suppose that at a high redshift of $z\geq100$, there exists a quasar-like object with a luminosity of about $10^{10}$ solar luminosities. The existence of such an object requires a formation mechanism beyond standard stellar evolution. As a specific example, we will consider the option of a PBH that emits radiation in a wide range of the spectrum as a result of accretion of the surrounding matter.

To illustrate the calculation method, we assume that the luminosity $L$ during accretion is equal to a fixed fraction $\lambda$ of the Eddington luminosity $L_{\rm Edd}$:
\begin{equation}
L=\lambda L_{\rm Edd}\simeq10^{44}\left(\frac{\lambda}{10^{-2}}\right)\left(\frac{M_{\rm PBH}}{10^8M_\odot}\right)\mbox{~erg~s$^{-1}$},
\end{equation}
where $M_{\rm PBH}$ in the PBH mass. Further, in the calculations, all assumptions about accretion at the PBH correspond to those made in \cite{Car81}, and the characteristics of radiation processes are calculated using the expressions given in monograph~\cite{ZelRai63}. In particular, following~\cite{Car81}, we assume a flat spectrum of radiation emitted during accretion:
\begin{equation}
L_\nu=\frac{dL}{d\nu}=\frac{L}{\nu_\text{max}}e^{-\nu/\nu_\text{max}},
\label{speq}
\end{equation}
where $\nu_{\max}=E_\text{max}/h$, $E_\text{max}\sim1$~keV--$1$~Mev, and $h=2\pi\hbar$ is the Planck constant. We assume that this spectrum extends to energies below 10~eV. In this case, the radiation from such an object may include a powerful Lyman-$\alpha$ emission.

The flatness of the spectrum down to a few eV is crucial for populating the $n=2,3,4$ levels. In a standard thin accretion disk, the temperature decreases with the radius as $T(r)\propto r^{-3/4}$, reaching a minimum $T_{\rm min}$ at the outer edge. If the disk extends to large radii (e.g., $r_{\rm out}\sim 10^5$--$10^6$ gravitational radii), $T_{\rm min}$ can be as low as $\sim$$0.1$--1 eV, providing the required few-eV photons. If the disk is truncated or $T_{\rm min}$ exceeds a few eV, the population enhancement is significantly reduced. Thus, the existence of a sufficiently extended accretion disk is essential for the proposed mechanism. In addition to the analytical calculations, there are also detailed numerical calculations of the accretion process and the resulting radiation spectrum. In~\cite{YuaNar14}, the numerical calculation of the spectra was performed for a set of black hole masses $M/M_\odot=10$, $10^3$, $10^5$, $10^7$, and $10^9$ (Figure 1 in~\cite{YuaNar14}). The results of these calculations for the minimum photon energy can be represented as the fitting formula $E_{\rm min}=(10M_\odot/M)^{1/2}$~eV~\cite{Pouetal17}. As can be seen from this expression, the minimum energies of $<$$1$~eV are easily achieved for massive PBHs. Recall that we are using a PBH as an example; the emission spectra of other massive objects may be completely different.

Under the influence of radiation with the spectrum (\ref{speq}), the hydrogen atoms near the PBH will be excited and even ionized in the very near zone. Consider the region at a distance of $r$ from the PBH. Let this distance be large enough so that the degree of ionization of the gas in the region under consideration is $\ll$$1$, and the majority of the hydrogen atoms are in the ground state with $n=1$. Then, an increase in the population of excited levels will occur in transitions $1\to n$ from this ground level. The number of transitions $1\to n$ per unit of time per unit volume is
\begin{equation}
\dot N_\text{in}=\frac{L_\nu}{4\pi r^2 h\nu}n_H(z)\int \sigma_{\nu1n}d\nu
\label{jin}
\end{equation}
where $\nu$ in this expression can be taken at $\nu=\nu_1=E_1/h$, $E_1=13.6$~eV; $n_H(z)$ is the number density of the neutral hydrogen. The cross-section integrated along the line profile has the form~\cite{ZelRai63}
\begin{equation}
\int \sigma_{\nu1n}d\nu=\frac{\pi e^2}{mc}f_{1n}\simeq2.64\cdot10^{-2}f_{1n}\mbox{~cm$^2$~s$^{-1}$},
\end{equation}
where $e$ and $m$ are the electron charge and mass, and the corresponding oscillator strength is $f_{1n}\simeq1.96/n^3$.

The decrease in the population of level $n$ under the influence of CMB radiation can be written as
\begin{equation}
	\dot N_\text{out}=\int\limits_{\tilde\nu_n}^{\infty}d\nu\sigma^\text{ex}_{\nu n}\frac{B_\nu}{h\nu}n_n(z),
	\label{jout}
\end{equation}
where
\begin{equation}
\sigma^\text{ex}_{\nu n}=\frac{64\pi^4}{3\sqrt{3}}\frac{e^{10}m}{h^6c\nu^3n^5},
\end{equation}
\begin{equation}
B_\nu=\frac{2h\nu^3}{c^2}\frac{1}{e^{h\nu/kT}-1},
\end{equation}

Here, $k$ is the Boltzmann's constant; $n_n(z)$ is the number density of the excited hydrogen atoms in the $n$ state; at the red-shift $z$, the CMB temperature is $T=2.7(1+z)$~K, $h\tilde\nu_n=E_1/n^2$. Electrons at highly excited levels behave quasi-classically with high accuracy; therefore, the expression (\ref{jout}) takes into account both transitions to higher levels and transitions to a continuous spectrum.

Let us denote
\begin{equation}
J(x)=\int\limits_{x}^\infty\frac{dx}{x\left(e^x-1\right)}.
\end{equation}

For $x\gg1$, one has $J(x)\simeq \Gamma[0,x]\simeq x^{-1}e^{-x}$, where $\Gamma[0,x]$ is the incomplete gamma function. 
By equating the two above expressions $\dot N_\text{in}=\dot N_\text{out}$, we get the relative population of the level:
\begin{equation}
\frac{n_n}{n_H}\simeq10^{-11}\left(\frac{L}{10^{44}\mbox{~erg~s$^{-1}$}}\right)\left(\frac{n}{4}\right)^2\left(\frac{r}{50\mbox{~kpc}}\right)^{-2}\left(\frac{J(h\tilde\nu_n/(kT))}{1.24\cdot10^{-12}}\right)^{-1}\left(\frac{E_\text{max}}{100\mbox{~keV}}\right)^{-1}.
\label{eq:pop}
\end{equation}

\begin{figure}[H]
	\includegraphics[width=0.8\textwidth]{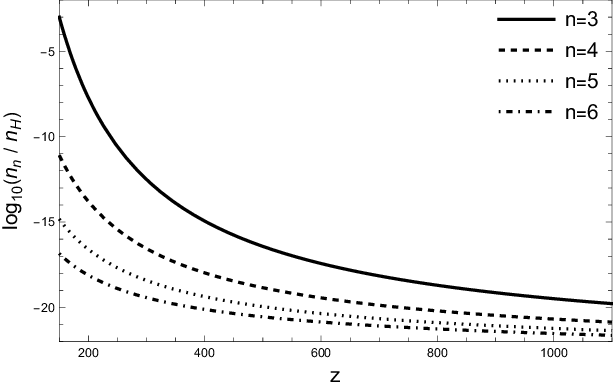}
	\caption{Populations of the hydrogen levels $n=$ 3, 4, 5 and 6 caused by PBH radiation depending on the redshift $z$.}
	\label{gr1}
\end{figure}

The normalization coefficients correspond to the epoch at $z=150$. The result of calculations for the same values of $r$, $L$, and $E_\text{max}$ are shown at Figure~\ref{gr1} for larger $z$.

In addition to process (\ref{jout}) discussed above, the spontaneous level decay and collisional deactivation also lead to the destruction of excited states of hydrogen atoms. The main spontaneous process is the transition to the first level with the rate
\begin{equation}
	A_{n,1}\simeq\frac{1.6\times10^{10}}{n^5}\mbox{~s$^{-1}$}.
\end{equation}

The next most important decays are $A_{n,2}=A_{n,n-1}=A_{n,1}/2$, while other transitions are less likely. Thus, the rate of spontaneous decays per unit volume is $\sim$$2A_{n,1}n_n$. The collisional deactivation cross-section for collisions of an excited atom with surrounding neutral atoms has the form $\sigma_{\rm coll}\pi r_B^2n^4$, where the Bohr radius is $r_B=\hbar^2/(e^2m_e^2)$. Therefore, the rate of collisional deactivation per unit volume is $\sim$$\sigma_{\rm coll}vn_nn_H$, where thermal velocity is $v\sim(k_BT/m_p)^{1/2}$. A comparison with (\ref{jout}) shows that (\ref{jout}) at $n>3$ is several orders of magnitude larger than these two additional processes at $z>150$. At $n=3$, the rate of spontaneous decay is compared with (\ref{jout}) only at $z\simeq150$, and at $z>150$, the rate (\ref{jout}) remains the largest. At the same time, the rate of collisional deactivation always remains subdominant. Thus, the expression (\ref{jout}) is valid in all cases of interest to us.

As a result, the populations of the excited states of hydrogen turns out to be many orders of magnitude larger than that given by the Boltzmann factor for the populations of the excited states of hydrogen in the pure CMB background. Really, the Boltzmann equilibrium (Saha equation) suggests
\begin{equation}
\frac{n_n}{n_H}\simeq\frac{g_n}{g_1}\exp\left\{-\frac{E_1}{kT}\left(1-\frac{1}{n^2}\right)\right\},
\label{boleq}
\end{equation}
where $g_n=2n^2$. The expression (\ref{boleq}) gives, for $n=4$, the value $n_n/n_H\sim10^{-167}$ at $z=150$ and $n_n/n_H\sim10^{-24}$ at $z=1000$. This significant increase in population means that the populations of the subordinate levels of the hydrogen atom also increase sharply.

\subsection{Radii and Gas Heating}

Two important scales in the problem under consideration are the stopping radius of the dark matter layer and the ionization radius. The stopping radius is the radius at which the cosmological expansion of the layer stopped, and the layers with smaller radii began to shrink until they eventually virialize when compressed by about half. The importance of this scale is in the fact that, within the stop radius, the density of dark matter is already noticeably higher than the average density. The stop radius is
\cite{Ber85}
\begin{equation}
	R_{\rm stop}(z)=0.12\left(\frac{M_{\rm PBH}}{10^8M_\odot}\right)^{1/3}\left(\frac{1+z}{501}\right)^{-4/3}~\mbox{kpc}.
	\label{rseq}
\end{equation}

The ionization radius is the radius within which hydrogen is almost completely ionized under the action of PBH radiation. This radius was calculated in the work~\cite{Car81} as
\begin{equation}
	R_{\rm ion}=\left(\frac{2L}{3\pi\alpha n_H^2 E_{\rm max}}\right)^{1/3},
	\label{rio1}
\end{equation}
where the ionization coefficient is
\begin{equation}
	\alpha=2.6\times10^{-13}\left(\frac{T}{10^4\mbox{K}}\right)^{-0.85}\mbox{~cm$^3$s$^{-1}$}.
	\label{alpeq}
\end{equation}

\begin{figure}[H]
	\includegraphics[width=0.8\textwidth]{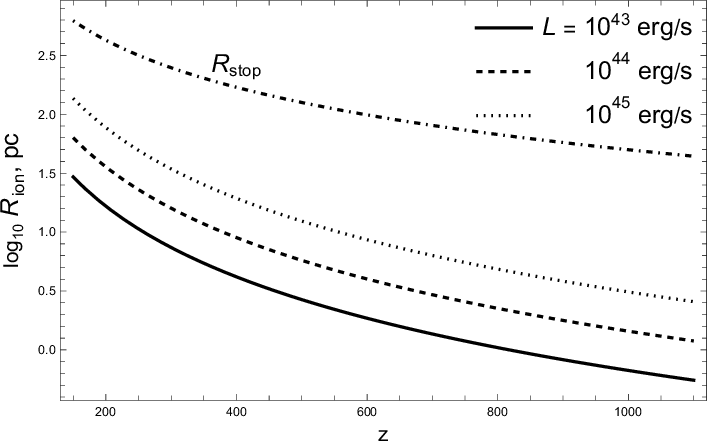}
	\caption{The ionization radius (\ref{rio1}) for the various luminosities, $L$, of the central object, as well as the stopping radius (\ref{rseq}) [dot-dashed curve] of the layer for PBH mass $10^8M_\odot$.}
	\label{gr2}
\end{figure}

These two scales are shown in Figure~\ref{gr2}, where the value of $n_H$ is taken near the stop radius $R_{\rm stop}$, which gives an upper estimate. It can be seen from the figure that the ionization radius is located inside the dark matter halo, which is formed around the PBH. The observation of subordinate lines is possible only in areas at distances $r>R_{\rm ion}$ from the PBH.

In assessing the prospects for the registration of hydrogen lines, it is necessary to discuss the increase in gas temperature due to heating by radiation from the PBH and heating during the compression of gas when its density increases near the PBH. However, at $z>150$, the reverse cooling process with Compton photon scattering is predominant; therefore, to a first approximation, the gas temperature can be considered equal to the temperature of the CMB. More precisely, baryons cool slightly faster, but the relative temperature difference is very small; at the redshift of $z\sim1000$, it is $\sim$$10^{-6}$ \cite{Kho11}. Therefore, it is difficult to observe absorption in lines at such large $z$. On the contrary, the possibility of observing the emission subordinate lines is present even at such large redshifts.

The central source of radiation, if it has a large mass, causes an increase in the density of dark matter and baryons around it. This increase could lead to an increase in the temperature of the gas. As mentioned above, at high $z$, the effect of gas cooling is due to Thomson scattering of CMB photons. However, as one approaches $z=150$, this process becomes ineffective, and an adiabatic increase in temperature during compression can be an important effect. In the work of \cite{Ber85}, the law of density increase towards the center was obtained in parametric form. We introduce the smallness parameter
\begin{equation}
\xi\approx\left(\frac{3\pi}{4}\right)^{4/3}\left(\frac{r}{R_{\rm stop}}\right)^{-3/2},
\end{equation}
and decompose the solution from \cite{Ber85} over this parameter. We find the asymptotics of the density profile in the form
\begin{equation}
\frac{\rho}{\bar\rho}\approx1+\frac{12}{175}\xi^4
\label{ser}
\end{equation}
at $\xi\ll 1$.
In monatomic gas, the adiabatic temperature increase occurs according to the law
\begin{equation}
\frac{\delta T}{T}=\frac{2}{3}\frac{\delta \rho}{\rho}\simeq4.4\left(\frac{r}{R_{\rm stop}}\right)^{-6}.
\end{equation}

Thus, at large distances from the object $r\gg R_{\rm stop}$, the gas density decreases very rapidly, and its temperature rise is negligible. At the same time, the population of the upper levels, according to (\ref{boleq}), decreases much more slowly as $r^{-2}$. In this paper, observations of the object from a cosmological distance at large $z$ are assumed; therefore, only the far peripheral region of gas around the central object (e.g., PBH), where the adiabatic temperature increase is very small, will be available for observations. Thus, gas heating is insignificant at both high and low $z$, and the effect under consideration has the prospect of being observed.

\subsection{Optical Depth Estimate and Validity of the Two-Level Approximation}
\label{sec:optdepth}

To substantiate the $\tau\sim 1$ claim, we explicitly calculate the optical depth in a subordinate line. The line's optical depth is given by
\begin{equation}
\tau_\nu = \int \sigma_\nu(\nu) \left(n_l - \frac{g_l}{g_u} n_u\right) dl,
\end{equation}
where $n_l$ and $n_u$ are the populations of the lower and upper levels, $g_l$ and $g_u$ are their statistical weights, and $\sigma_\nu(\nu)$ is the normalized line profile. For a thermal gas, the profile is Doppler-broadened with width
\begin{equation}
\Delta\nu_D = \nu_0 \sqrt{\frac{2k_B T}{m_H c^2}}.
\end{equation}

At $z=150$, gas is thermally coupled to the CMB, so $T\simeq T_{\rm CMB}=2.7(1+z)\simeq 400$ K. Substituting the constants $k_B = 1.38\times10^{-16}$ erg K$^{-1}$, $m_H = 1.67\times10^{-24}$ g, and $c = 3.0\times10^{10}$ cm s$^{-1}$, we obtain $\Delta\nu_D\simeq 10^{-5}\nu_0$. Turbulent motions in the halo may increase this to $\sim10^{-4}$--$10^{-3}$ depending on the local velocity field.

For the $5\to6$ transition at $z=150$, using the population ratio $n_n/n_H\sim10^{-11}$ from Equation~(\ref{eq:pop}), the hydrogen density $n_H(z=150)\simeq 2\times10^{-1}\,\text{cm}^{-3}$, and a path length $L\sim 50$~kpc (the characteristic scale of the enhanced-density halo), we obtain a column density of excited atoms $N_{\rm col} = n_n L \simeq 3\times10^{11}\,\text{cm}^{-2}$. Using the integrated cross-section $\int \sigma_\nu d\nu = (\pi e^2/mc) f_{ln}$ with oscillator strength $f_{5\to6}\simeq 0.14$ and the Doppler width $\Delta\nu_D \simeq 10^{-5}\nu_0$, we estimate the peak optical depth as
\begin{equation}
\tau \sim \frac{\int \sigma_\nu d\nu}{\Delta\nu_D} N_{\rm col} \sim \frac{(\pi e^2/mc) f_{5\to6}}{\Delta\nu_D} N_{\rm col}.
\end{equation}

Numerically, $(\pi e^2/mc) \simeq 2.65\times10^{-2}$ cm$^2$ s$^{-1}$; for $\nu_0 \simeq 4.0\times10^{13}$ Hz (rest-frame frequency of the $5\to6$ transition), $\Delta\nu_D \simeq 4.0\times10^8$ Hz. Thus, $\tau \sim 3$. This confirms that $\tau\sim1$ (or greater) is achievable in the halo, provided that the gas density and population enhancement are sustained over the path length. We emphasize that the geometry is not a simple slab: The emission originates from a spherical shell or ring at radii where $r\gg R_{\rm ion}$ and $n_n\propto r^{-2}$, so the effective path length is set by the radial extent of the enhanced-population region, which can be tens of kiloparsecs. In a more precise approach, one can consider the passage of a photon at an impact distance $L$ from the center of the object. We use the angular variable $\phi$, and we write $r=L/\cos\phi$ and $s=L\tan\phi$. Integrating over $\phi$ from $-\pi/2$ to $\pi/2$, in the case of $n_n\propto r^{-2}$, we get $\int n_n(r)ds=\pi L n_n(L)$. With turbulent broadening of $\sim10^{-4}$, $\tau$ would be $\sim$$0.3$, which is still of order unity. Thus, regardless of the exact broadening mechanism, $\tau\sim1$ is a robust result.

It is important to clarify which physical processes are neglected in Equation~(\ref{eq:pop}) and why they are subdominant. We have intentionally used a simplified two-level description to obtain an order-of-magnitude estimate. The following processes are omitted:
\begin{itemize}
\item Spontaneous Radiative Decay $n\to n'$: At high $n$, the Einstein $A$-coefficients scale as $A_{ul}\propto n^{-5}$, making spontaneous decay very slow compared to the CMB-driven upward transitions. For $n\sim5$--$6$, $A_{ul}\sim10^6\,\text{s}^{-1}$, while the CMB excitation rate is $\sim$$10^8\,\text{s}^{-1}$, so the latter dominates.
\item Collisional Processes: At $z\sim150$, the gas density is $n_H\sim10^{-1}\,\text{cm}^{-3}$, and the collisional rate is $n_H \langle \sigma v \rangle \sim 10^{-9}\,\text{s}^{-1}$, far below the radiative rates.
\item Lyman-$\alpha$ Radiative Transfer: While multiple scattering of Lyman-$\alpha$ photons can affect the ground-state population, it does not significantly alter the high-$n$ populations because the high-$n$ levels are populated directly by the continuum radiation from the central source.
\item Two-Photon Decay from $2s$: This process affects the $2s$ level, but our focus is on $n\ge3$ levels, which are populated directly and are not coupled to $2s$ on the relevant timescales.
\item Lyman-$\alpha$ Escape in an Expanding Medium: The Universe at $z\sim150$ is expanding, but the physical scale of the halo ($\sim$pc--kpc) is much smaller than the Hubble scale, so expansion does not significantly alter the optical depth within the halo.
\end{itemize}

Thus, Equation~(\ref{eq:pop}) provides a robust order-of-magnitude estimate, and the neglected processes would introduce corrections of order unity at most.

At a distance of $\sim$$50$~kpc from the PBH, gas is predominantly neutral, and its equation of state can be approximated as ideal: $P = n k_B T$, where $n$ is the number density of hydrogen atoms and $T$ is the gas temperature. Using the density profile from Equation~(15) and the CMB temperature (since at $z\gtrsim150$, the gas is thermally coupled to the CMB), we estimate values for $z\sim150$: $n\sim10^{-2}\,\text{cm}^{-3}$, $T\sim 400$, and hence, $P\sim10^{-16}\,\text{dyn\,cm}^{-2}$.

The recombination rate in this region is $\dot n_{\rm rec} = \alpha_B n_e n_p$. For case B's recombination at $T\sim400$ K, the coefficient is $\alpha_B \simeq 4.0\times10^{-12}\,\text{cm}^3\,\text{s}^{-1}$ (using the scaling $\alpha_B \propto T^{-0.85}$ from~\cite{Car81}). Assuming full ionization ($n_e\simeq n_p\simeq n_H\simeq10^{-2}\,\text{cm}^{-3}$), we get $\dot n_{\rm rec}\sim4\times10^{-16}\,\text{cm}^{-3}\,\text{s}^{-1}$. In reality, the gas is mostly neutral, so $n_e$ and $n_p$ are much smaller (the residual ionization of the Universe is $3.1\times10^{-4}$), making $\dot n_{\rm rec}$ even lower. This is far below the excitation rate $\dot N_\text{in}$, confirming that the population is driven by external radiation.


\section{Prospects for Observing the High-\emph{n} Absorption Lines}

\subsection{Observational Advantages of High-\emph{n} Transitions}

Recall that the populations of the high-$n$ levels, while remaining in absolute values that are significantly less than unity (on the order of $10^{-11}$), are nevertheless $\sim$$10$ orders of magnitude larger than they would be in the equilibrium case. This allows one to achieve an optical thickness in the subordinate lines on the order of unity ($\tau \sim 1$) in a halo around a PBH.

Such an optical thickness makes it possible to observe lines in a manner similar to that described for the 21\,cm line in~\cite{Dubrovich2021b} due to another important effect---the temperature difference between matter and radiation. If we observe radiation in subordinate lines at $z\sim 150$, the temperature difference can be on the order of the temperature itself, i.e., the gas temperature can be several tens of percent lower than the CMB temperature.

Thus, we have the opportunity to generate noticeable Fraunhofer-like absorption in the aforementioned subordinate lines within the sufficiently distant atmosphere surrounding a possible massive object. The observational manifestations appear in a straightforward and obvious manner: by observing the photosphere of the halo of this object, we see absorption lines corresponding to subordinate transitions in the hydrogen atom.

Probably, redshifts of $z\sim200$--$300$ are most suitable for observing subordinate lines. In this case, the difference between the temperature of hydrogen and the temperature of cosmic microwave background radiation is already quite large. At the same time, the large redshift transfers the lines into the radio band and makes them available for observations using radio telescopes.

The frequencies of these lines are several times, almost an order of magnitude, different from the Lyman-$\alpha$ frequency. Today, due to large redshifts, these lines fall into the range of radio emission. This is a fundamentally important effect. When observing the Lyman line as initially emitted at $z \approx 1000$, this line would have an observed wavelength of approximately $100\,\mu$m, which is inaccessible to ground-based telescopes. Meanwhile, the subordinate lines will fall into the radio range. This range is accessible to a large number of modern telescopes. The frequency separation between the subordinate lines in this spectral region is about $30\text{--}40\%$, which fits well within the spectral window of modern radio telescope systems.

Regarding the detection of Balmer and Paschen lines at $z\sim100$, as observed in~\cite{Zhaetal26} at $z=4.5$, these lines (transitions with $n=2,3$) at $z\sim100$ would be redshifted to far-infrared wavelengths ($\lambda_{\rm obs}\sim 100$--$300\,\mu$m), which are observable only from space-based telescopes. In contrast, our proposed method focuses on transitions with $n\ge5$, which fall into the radio band and are accessible from ground-based radio telescopes. Thus, while Balmer/Paschen lines at $z\sim100$ are challenging, our approach offers a complementary and more feasible observational route.

Regarding line broadening, the dominant mechanism is Doppler broadening due to thermal motions: $\Delta\nu_D/\nu \sim (2k_BT/m_H c^2)^{1/2} \sim 10^{-5}$ at $T\sim400$ K. Pressure (collisional) broadening is negligible: $\Delta\nu_{\rm col} \sim n \sigma v \sim 10^{-9}$ Hz, far smaller than the Doppler width.

The key observational advantage of subordinate transitions (e.g., Balmer, Paschen, Brackett) over the resonant Lyman-$\alpha$ line is that their rest-frame wavelength $\lambda_{\text{rest}}$ is longer by factors of $n^2$ (for high principal quantum numbers). In contrast, Lyman-$\alpha$ at $\lambda_{\text{rest}} = 121.6\ \text{nm}$ is redshifted to far-infrared for $z\sim 1000$:

\[
\lambda_{\text{obs}} = \lambda_{\text{rest}} \cdot (1+z),
\]

\[
\lambda_{\text{obs}}(\text{Ly-}\alpha) = 121.6\ (1+z)\ \text{nm} \approx 122\ \mu\text{m}\ (z=1000),
\]
which is blocked by the Earth's atmosphere and requires space telescopes. In contrast, a Balmer line like H-$\alpha$ ($\lambda_{\text{rest}}=656.3\ \text{nm}$) or a Paschen line ($\lambda_{\text{rest}}=1875\ \text{nm}$) gives
\[
\lambda_{\text{obs}}(\text{H-}\alpha) = 656.3\ (1+z)\ \text{nm} \approx 0.66\ \text{mm}\ (z=1000),
\]
\[
\lambda_{\text{obs}}(\text{Paschen-}\alpha) = 1875\ (1+z)\ \text{nm} \approx 1.9\ \text{mm}\ (z=1000).
\]

For $z\sim 300$, these shift to micrometre waves ($\sim$$200\ \mu\text{m}$ for H-$\alpha$; $\sim$$600\ \mu\text{m}$ for Paschen-$\alpha$), perfectly matching ground-based radio telescopes, such as the upgraded Giant Metrewave Radio Telescope (uGMRT), the Square Kilometre Array (SKA-Low), and space-based radio telescopes such as planned Spektr-M. Moreover, resonant Lyman lines are optically thick in the halo and suffer from strong absorption by neutral hydrogen along the line of sight, whereas subordinate transitions are optically thinner ($\tau\sim 1$ under the proposed population enhancement) and directly reveal the temperature difference between gas and CMB as Fraunhofer-like absorption. Thus, subordinate lines turn an otherwise invisible far-IR object into a detectable radio source.

Let us consider the transitions between high principal quantum numbers ($n \sim 6$). Besides the Balmer, Paschen, and Brackett lines, transitions between highly excited states of the hydrogen atom, such as $5 \rightarrow 6$ and $6 \rightarrow 7$, are of particular interest. Their rest-frame wavelengths are given by the Rydberg formula~\cite{Biraben2009}:

\[
\frac{1}{\lambda_{\text{rest}}} = R_{\text{H}} \left( \frac{1}{n_{\text{low}}^2} - \frac{1}{n_{\text{high}}^2} \right),
\]
where $R_{\text{H}} \approx 1.097 \times 10^7\ \text{m}^{-1}$. For the selected transitions, we obtain

\begin{itemize}
    \item Transition $5 \rightarrow 6$: $\lambda_{\text{rest}}(5\to6) \approx 7.46\ \mu\text{m}$.
    \item Transition $6 \rightarrow 7$: $\lambda_{\text{rest}}(6\to7) \approx 12.36\ \mu\text{m}$.
\end{itemize}

At high redshifts $z \sim 300$ and $z \sim 1000$, these lines are redshifted into the millimetre and centimetre bands:

For $z = 300$,
\[
\lambda_{\text{obs}}(5\to6) \approx 2.24\ \text{mm}, \qquad
\lambda_{\text{obs}}(6\to7) \approx 3.72\ \text{mm}.
\]

For $z = 1000$,
\[
\lambda_{\text{obs}}(5\to6) \approx 7.47\ \text{mm}, \qquad
\lambda_{\text{obs}}(6\to7) \approx 12.37\ \text{mm}\ (1.24\ \text{cm}).
\]

Thus, at $z \sim 1000$, these transitions fall into the centimeter range ($1\text{--}2\ \text{cm}$), accessible to ground-based radio telescopes (SKA, uGMRT, and VLA). At $z \sim 300$, they lie in the millimetre range ($2\text{--}4\ \text{mm}$) and are also observable from Earth under favorable atmospheric conditions (ALMA and NOEMA) and from space (Spektr-M).

Such transitions between closely spaced high levels ($\Delta n = 1$) have very small Einstein coefficients ($A_{ul} \propto n^{-5}$), making them extremely weak under equilibrium conditions. However, as shown in Section~\ref{sec:mechanism}, the Lyman-$\alpha$ radiation from a primordial quasi-stellar object boosts the populations of high levels by $8\text{--}10$ orders of magnitude, making these transitions optically thick ($\tau \sim 1$) in a halo of tens of kiloparsecs. Hence, observing the $5\to6$ and $6\to7$ lines (and also higher transitions, e.g., $6\to8$) provides a unique tool for probing primordial compact objects during the Dark Ages.

Unlike the resonant Lyman-$\alpha$ line, which remains in the far-IR for $z>6$ and is inaccessible from the ground, subordinate transitions with $n \sim 6$ yield a clear spectral fingerprint in the radio band. Their relative frequency separation ($\Delta \nu / \nu \sim 0.3$ for $5\to6$ and $6\to7$) is well within the resolving power of modern radio spectrometers.

\subsection{Key Observational Parameters}\label{sec3.2}

\begin{figure}[H]
	\includegraphics[width=0.8\textwidth]{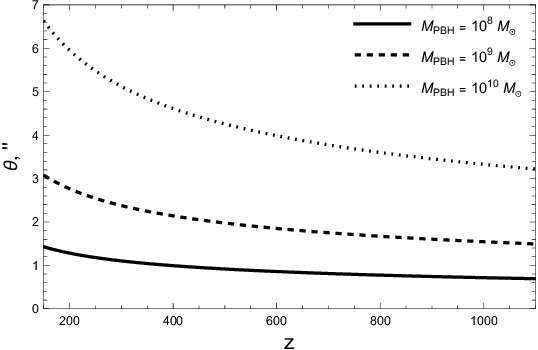}
	\caption{The angular size of the stop radius $R_{\rm stop}$, Equation~(\ref{rseq}), depending on the redshift for the PBH masses $10^8M_\odot$, $10^9M_\odot$, and $10^{10}M_\odot$ (from bottom to top). It is proposed to search for the subordinate lines for the areas $r\gg R_{\rm stop}$ around the PBHs. Accordingly, the angular size of the observed area is about 1--2 orders of magnitude larger than shown in this figure.}
	\label{gr3}
\end{figure}

We now derive the three key observational numbers:
\begin{itemize}
\item Line Depth ($\sim$$10\%$ of Continuum): The line depth is given by $\Delta T/T \simeq \tau (T_{\rm CMB}-T_{\rm gas})/T_{\rm CMB}$. With $\tau\sim1$ and $T_{\rm gas}\simeq 0.9 T_{\rm CMB}$ at $z\sim150$, we obtain $\Delta T/T \sim 0.1$, i.e., a temperature variation of $\Delta T \sim 0.1 \times 2.7(1+z) \sim 0.1 \times 400 \sim 40$ K in the rest frame or $\sim$$1$ K in the observed frame after redshift dilution. This is consistent with the $10\%$ estimate.
\item Relative Line Width ($10^{-3}$--$10^{-4}$): The line width is dominated by Doppler broadening: $$\Delta\nu/\nu \sim (2k_B T_{\rm gas}/m_H c^2)^{1/2} \sim 10^{-5}$$ for $T\sim400$ K. Turbulent broadening could increase this to $\sim$$10^{-3}$, hence the quoted range.
Stark broadening can be estimated using the simple formula $\Delta\nu\sim8.2n_e(n/100)^{4.5}$, where the electron number density is $n_e=xn_H$ \cite{GorSor09}. For the physical conditions considered in this paper (at $z\geq150$, $x\simeq3.1\times10^{-4}$), $\Delta\nu\ll1$ is obtained, so the effect of Stark broadening does not play a role.
\item Angular Size ($\sim$$10$ Arcseconds): This is the physical size of the halo where $\tau\sim1$ is $R\sim 50$~kpc (the stop radius). At $z\sim150$, the angular diameter distance is $D_A\sim 1.5$~Gpc, so $\theta \sim R/D_A \sim 50\,\text{kpc}/1.5\,\text{Gpc} \sim 3\times10^{-5}\,\text{rad} \sim 6$ arcseconds. Accounting for a larger halo or a somewhat lower $z$ gives $\sim10$ arcseconds.
A well-known effect in cosmology is that the angular distance, unlike the photometric distance, not only does not grow at large redshifts but even decreases slowly with the redshift. This leads to the fact that the angular size of distant cosmological objects at $z>1.5$ remains large, and from the point of view of the angular resolution of telescopes, their observation is not a problem. Interestingly, the increase in the angular size almost compensates for the decrease in the stopping radius (\ref{rseq}) with the redshift (see Figure~\ref{gr3}).
\end{itemize}

\subsection{Sensitivity Estimate for Future Radio Telescopes}

To make the observability claim quantitative, we estimate the integration time required to detect a $10\%$ absorption line with a modern radio telescope. We use the following radiometer equation:
\begin{equation}
\sigma_T = \frac{T_{\rm sys}}{\sqrt{\Delta \nu \, t_{\rm int}}},
\end{equation}
where $T_{\rm sys}\sim 100$~K is the system's temperature (typical for SKA mid-band), $\Delta \nu$ is the bandwidth, and $t_{\rm int}$ is the integration time. For a line with $\Delta \nu/\nu \sim 10^{-3}$ at $\nu\sim 1$~GHz (corresponding to $z\sim 1000$ for a $5\to6$ transition), $\Delta \nu \sim 1$~MHz. To achieve a signal-to-noise ratio of $5$ for a $10\%$ ($0.1\times T_{\rm sys}$) line, we need $\sigma_T \lesssim 0.02 T_{\rm sys}$, which gives
\begin{equation}
t_{\rm int} \gtrsim \frac{1}{\Delta \nu} \left(\frac{T_{\rm sys}}{0.02 T_{\rm sys}}\right)^2 \approx 2.5\times10^{-3}\,\text{s}.
\end{equation}

This integration time is very short, indicating that the line itself is bright. However, the beam's dilution must be considered: A $10''$ source at 1~GHz has a beam size of $\sim$$1'$ for SKA, so the source fills only a small fraction of the beam. The effective system temperature is increased by the beam dilution factor $\sim (1'/10'')^2 \sim 36$, giving $T_{\rm sys}^{\rm eff} \sim 3600$~K and then $t_{\rm int} \gtrsim 2.5\,\text{s}$. Thus, even with beam dilution, the integration time is of order seconds. For weaker lines or smaller sources, the integration time scales as $(\theta_{\rm source}/\theta_{\rm beam})^4$ and could reach hours. This demonstrates that the proposed effect is within the reach of current and planned radio telescopes such as SKA and uGMRT.


\section{Discussion}

\subsection{Little Red Dots}

Recently, the JWST has revealed a population of compact red objects at $z\approx4$–$10$ known as ``Little Red Dots'' (LRDs) \cite{Zhang2026, Taylor2025}. These are interpreted as obscured AGN with massive black holes ($10^6$–$10^8\,M_\odot$) and angular sizes of $\sim 0.1''$–$1''$. While their redshifts are lower than those considered here, they demonstrate that massive black holes exist at early epochs. These black holes can be primordial, and they can be surrounded by dense gas~\cite{Zhaetal26}. Extrapolating their properties to higher redshifts suggests that similar objects could be observable via the proposed absorption-line method.

The predicted enhancement of level populations (see Section~\ref{sec:mechanism}) provides the necessary optical depth for detectable absorption. The main novelty is the prediction of detectable absorption features in the radio band (accessible to the near-future radio telescopes) rather than in the far-IR, providing a practical observational avenue. Very high spectral resolutions are not required, and thus, no special efforts are needed to obtain results. The sensitivity can be standard, with a noise temperature of around $100\,$K, which allows one to accumulate the necessary signal in a reasonable amount of time. The number of such sources in the sky cannot be precisely determined at this time.

We note that the obscuration of LRDs is not merely a contextual detail; it is precisely what makes the proposed method attractive compared to conventional techniques. As shown by~\cite{Peca2025a,Zhang2026}, high-$z$ sources are predicted to be obscured, and thus, they are accessible only with future facilities. If the obscuring material is largely neutral hydrogen, the same medium that hides the central engine is the one in which the excitation of subordinate transitions occurs. Thus, the proposed method is most effective precisely in the conditions that make such objects challenging to observe by traditional means, strengthening the motivation for this work. We refer the reader to recent reviews on obscured accretion at high redshifts~\cite{Peca2025a, Peca2025b} and the references therein for further discussion.

\subsection{Possible Competing Signals}

Two additional effects may contaminate or compete with the proposed absorption-line signal:
\begin{itemize}
\item Thermal Sunyaev-Zel'dovich (SZ) Effect: The ionized region inside $R_{\rm ion}$ (see Equation~(\ref{rio1})) will produce a thermal SZ effect (Compton-$y$ distortion) on the CMB. The Compton parameter is $$y = \int (k_B T_e/m_e c^2) n_e \sigma_T dl$$. For $T_e\sim10^4$ K, $n_e\sim10^{-2}$ cm$^{-3}$, and for a path length of $\sim$kpc, we estimate $y\sim10^{-8}$--$10^{-7}$, which is below the current detection limit ($y\sim10^{-5}$) for individual clusters. Thus, the SZ effect is negligible compared to the line absorption.
\item Free--Free Emission: The H II region inside $R_{\rm ion}$ will produce free--free (bremsstrahlung) emission. The free--free optical depth is $$\tau_{\rm ff} \sim 0.1 (T_e/10^4\,\text{K})^{-1.35} (\nu/1\,\text{GHz})^{-2.1} (\text{EM}/10^{50}\,\text{cm}^{-5})\, ,$$where EM is the emission measure. For typical parameters, $\tau_{\rm ff}\ll1$ at GHz frequencies, so the free--free continuum is weak and would not significantly fill in the absorption line.
\end{itemize}

Thus, neither the SZ effect nor free--free emission is expected to contaminate the proposed absorption-line signal.

In radio astronomy, the transition lines between large $n$ levels are called ``recombination radio lines'' (RRLs), since the population of corresponding levels occurs during recombination in H I regions. In the case under consideration, the population of large $n$ levels is caused by excitation from external sources of radiation. In this regard, we can introduce the new term ``excitation radio lines'' (ERLs).

The main observable signatures are summarized in Section~\ref{sec3.2}: line depth of $\sim10\%$, relative width of $10^{-3}$--$10^{-4}$, and angular size of $\sim10$ arcseconds. These estimates are highly dependent on the chosen source model and should be regarded as rough, approximate values. However, the range of variation of these quantities allows us to hope that their detection is feasible.

\subsection{Plausibility of the Source Population}

The assumed $10^{10}L_\odot$ object at $z\ge100$ requires a formation mechanism beyond standard stellar evolution. Here, we briefly assess the plausibility of such objects:
\begin{itemize}
\item Number Density: If such PBHs constitute a fraction $f_{\rm PBH}$ of dark matter, their comoving number density is $n_{\rm PBH} = f_{\rm PBH} \rho_{\rm DM}/M_{\rm PBH}$. For $M_{\rm PBH}=10^8M_\odot$ and $f_{\rm PBH}=10^{-6}$, $n_{\rm PBH}\sim 10^{-10}\,\text{Mpc}^{-3}$, corresponding to $\sim$$10^3$ objects in the observable Universe. This is consistent with the rarity of luminous quasars at $z\sim6$--$7$.
\item Observational Constraints: PBHs with $M_{\rm PBH}\sim10^8M_\odot$ are constrained by CMB spectral distortions and accretion limits. However, as shown in~\cite{Ziparo2025, Hooper2024}, a non-Gaussian primordial power spectrum or clustering can evade these constraints. The assumed accretion rate $\lambda=10^{-2}$ is sub-Eddington and is stable at $z\sim150$ given the available gas supply within the dark matter halo (see Figure~\ref{gr2}).
\item Consistency with Observations: The proposed population is consistent with the lack of observed high-$z$ quasars, as the absorption-line signature would require dedicated radio observations to be detected.
\end{itemize}

Thus, while exotic, the required PBH population is not ruled out by current constraints.

\subsection{Future Prospects}

To assess observational feasibility, we estimate the CMB continuum flux in a $10''$ beam. At $\lambda=1$ cm, the specific intensity of the CMB is $I_\nu = 5.71\times10^7\,\text{Jy sr}^{-1}$. The solid angle of a $10''$ beam is $\Omega \approx \pi (5'')^2 = \pi (5/3600 \,\text{rad})^2 \approx 1.85\times10^{-9}\,\text{sr}$. Hence, the flux is $F_{\rm CMB} \simeq I_\nu \Omega \approx 0.106\,\text{Jy} = 106\,\text{mJy}$. This provides a reference level against which the line absorption depth of $\sim$$10\%$ (i.e., $\sim$$10$ mJy) should be detectable with modern radio telescopes.

The discovery of quasars at $z\sim100$ would be of extraordinary scientific significance, as it would probe the era during which the first compact objects formed, well before the epoch of reionization. It would provide direct evidence for the existence of primordial black holes or other exotic objects and would strongly constrain models of structural formation and the early thermal history of the Universe. To make precise predictions, future work should include full radiative transfer calculations coupled with realistic accretion disk models, taking into account the angular momentum distribution, disk structure, and frequency-dependent opacities. Such modeling will be essential to interpret future observations and to distinguish between different primordial object scenarios.

\vspace{6pt}
\noindent\textbf{Author Contributions:} All authors contributed equally to the conceptualization, methodology, and writing of this manuscript. All authors have read and agreed to the published version of this manuscript.

\noindent\textbf{Funding:} This research received no external funding.

\noindent\textbf{Data Availability Statement:} No new data were created or analyzed in this study. Data sharing is not applicable to this article.

\noindent\textbf{Acknowledgments:} The authors are grateful to the anonymous reviewers for their constructive comments and suggestions, which significantly improved the content of this article. The work of V.K. Dubrovich and S.I. Shirokov was performed as part of the State assignment of the Special Astrophysical Observatory of the Russian Academy of Sciences (approved by the Ministry of Science and Higher Education of the Russian Federation). The work of Yu.N. Eroshenko was performed as part of the State assignment of the Institute for Nuclear Research of the Russian Academy of Sciences.

\noindent\textbf{Conflicts of Interest:} The authors declare no conflicts of interest.

\noindent\textbf{Abbreviations:}

\begin{tabular}{@{}ll}
CMB & Cosmic microwave background \\
JWST & James Webb Space Telescope \\
PBH & Primordial black hole \\
SKA & Square Kilometre Array \\
uGMRT & Upgraded Giant Metrewave Radio Telescope \\
RRL & Recombination radio lines \\
ERL & Excitation radio lines
\end{tabular}

\end{document}